\documentclass[twocolumn,floatfix, showpacs]{revtex4}

\usepackage[final]{epsfig}
\usepackage{amsmath}
\usepackage{parskip}

\begin{document}

\title{Towards jet tomography: $\gamma$-hadron correlations}

\author{Thorsten Renk}
\email{trenk@phys.jyu.fi}
\affiliation{Department of Physics, PO Box 35 FIN-40014 University of Jyv\"askyl\"a, Finland}
\affiliation{Helsinki Institut of Physics, PO Box 64 FIN-00014, University of Helsinki, Finland}


\begin{abstract}
Hard pQCD processes taking place in ultrarelativistic heavy-ion collisions are a well-calibrated probe. It is believed that the interaction with the surrounding medium of outgoing partons from a hard vertex is capable of revealing details of the medium. We demonstrate that correlation measurements of hard photon-hadron back-to-back coincidences are a tool suitable to extract such tomographic information.
 Introducing the concept of averaged evergy loss probability distributions, we first argue that almost no information about details of the medium evolution is reflected in the nuclear suppression factor $R_{AA}$. Thus, a wide variety of jet quenching scenarios and geometries are compatible with the measured data. This problem can be overcome by a $\gamma$-hadron correlation measurement. We show that averaged probability distributions for quarks are accessible experimentally and sketch an analysis procedure capable of distinguishing different energy loss scenarios leading to the same nuclear suppression factor.

\end{abstract}

\maketitle

\section{Introduction}
\label{sec_introduction}

The expression 'jet tomography' often used to describe the analysis of hard pQCD processes taking place inside the soft matter created in an ultrarelativistic heavy-ion collision. In particular the focus is on the nuclear suppression of hard hadrons in A-A collisions compared with the scaled expectation from p-p collisions due to loss of energy from the hard parton by interactions with the soft medium (see e.g. \cite{Tomo1,Tomo2,Tomo3,Tomo4,Tomo5}). 

%The Oxford English Dictionary defines 'tomography' as 'Radiography in which 
%an image of a predetermined plane in the body or other object is obtained by 
%rotating the detector and the source of radiation in such a way that points 
%outside the plane give a blurred image. Also in extended use, any analogous 
%technique using other forms of radiation.'

The idea of jet tomography suggests that it is possible to make a 'cut' (greek $\tau \acute{o} \mu o \varsigma$) through the outer layers of the medium and study directly the hot and dense core of the early evolution of the fireball created in the collision process. 

However, the nuclear suppression factor 

\begin{equation}
R_{AA}(p_T,y) = \frac{d^2N^{AA}/dp_Tdy}{T_{AA}({\bf b}) d^2 \sigma^{NN}/dp_Tdy}.
\end{equation}

is a rather integral quantity, arising in model calculations from the schematical convolution of the hard pQCD vacuum cross section $d\sigma_{vac}^{AA \rightarrow f +X}$ for the production of a  parton $f$, the energy loss probability $P_f(\Delta E)$ given the vertex position and path through the medium and the vacuum fragmentation function $D_{f \rightarrow h}^{vac}(z, \mu_F^2)$ as 

\begin{equation}
\label{E-Folding}
d\sigma_{med}^{AA\rightarrow h+X} = \sum_f d\sigma_{vac}^{AA \rightarrow f +X} \otimes P_f(\Delta E) \otimes
D_{f \rightarrow h}^{vac}(z, \mu_F^2),
\end{equation}

where

\begin{equation}
d\sigma_{vac}^{AA \rightarrow f +X} = \sum_{ijk} f_{i/A}(x_1,Q^2) \otimes f_{j/A}(x_2, Q^2) \otimes \hat{\sigma}_{ij 
\rightarrow f+k}.
\end{equation}

Here, $f_{i/A}(x, Q^2)$ denotes the distribution of parton $i$ inside the nucleus as a function of the parton
momentum fraction $x$ and the hard scale $Q^2$ of the scattering and $\hat{\sigma}_{ij\rightarrow f+k}$ is the the partonic pQCD cross section \cite{pQCD-Xsec}.

As this expression has to be averaged over all possible vertex positions and paths for the emerging parton $f$, the question to what degree information about the structure of the medium is preserved and if such a measurement can be called tomography seems reasonable.

In this paper, the question is investigated as follows: First, the concept of the averaged energy loss probability distribution $\langle P(\Delta E, E) \rangle$ is introduced. We argue that this is a quantity which can be calculated in a wide variety of models (and is thus suitable for model comparison) and is (for quarks) also experimentally accessible. We demonstrate that a wide range of different $\langle P(\Delta E, E) \rangle$, corresponding to widely different physics mechanisms for jet quenching, yields a nuclear suppression factor compatible with the measured data. Thus, it can be argued that there is little tomogrphic information in $R_{AA}$. We show that the same variety of scenarios can easily be distinguished by measuring the spectrum of hard leading away side hadrons correlated with a high $p_T$ photon trigger and show to what degree tomographic information about the medium can be recovered from such a measurement.

\section{Averaged energy loss probability distributions}
 
Considering the structure of Eq.~(\ref{E-Folding}), one observes that all information about the medium is contained in the energy loss probability $P_f(\Delta E)$ for a given parton $f$ (while it is also possible to represent the energy loss of the leading hadron in the form of a medium-modified fragmentation function $D^{med}_{f\rightarrow h}(z, \mu_F^2)$, cf. e.g. \cite{QuenchingWeights} it is more convenient to retain the representation (\ref{E-Folding}) for the present discussion). In a less schematic representation, the evaluation of Eq.~(\ref{E-Folding}) hence requires spatial integrals over the possible vertex positions, angular integrals over the direction of propagation through the medium and line integrals along the path of the outgoing parton $f$ through the medium (see e.g. \cite{JetFlow,Fragility} for details). However, these spatial integrals factorize with the momentum-space ingredients $d\sigma_{vac}^{AA \rightarrow f +X}$ and $D_{f \rightarrow h}^{vac}(z, \mu_F^2)$. Thus, given a model for the medium and the energy loss, the integrals over all unobserved spatial quantities can be carried out separately, defining an averaged energy loss probability distribution $\langle P(\Delta E, E)\rangle$. This quantity can in principle retain dependence on the initial hard parton energy $E$ and contains all information about the medium convoluted with the energy loss model.

Such a quantity is also a highly intuitive way to characterize the medium: Given all assumptions about the medium and the energy loss, $\langle P(\Delta E, E)\rangle$ immediately reveals which fraction of partons is completely absorbed, which fraction does not undergo energy loss at all and which fraction is shifted by some $\Delta E$ in the spectrum before fragmentation occurs. In general, $\langle P(\Delta E, E)\rangle$ will be probabilistic --- either because the energy loss for a given parton path is treated on a probabilistic basis \cite{QuenchingWeights} or, if this is neglected, because the distribution of in-medium pathlength needs to be taken into account and the initial vertex and path of an individual parton is not observed \cite{Dainese,Tomo5}. 

On the other hand, the averaged probability distribution for quarks is in principle accessible experimentally using the $qg \rightarrow q\gamma$ process \cite{XNPhotons1,XNPhotons}. If a photon with a large transverse momentum $p_T^\gamma$ is used as trigger, the momentum spectrum of correlated high $p_T^h$ hadrons at angle $\approx \pi$ can be measured. The momentum spectrum can be calculated as
\begin{equation}
\begin{split}
\label{E-Analysis}
p_T^\gamma\frac{dN}{dp_T^h} =& \int_{z_{min}}^1 \negthickspace  \negthickspace \negthickspace \negthickspace
dz D^{vac}(z,Q^2) \int_0^1  \negthickspace
df \langle P(f, p_T^\gamma)\rangle  \delta\left(zf - \frac{p_T^h}{p_T^\gamma}\right) \\
& = \int_0^1 df D^{vac}(p_T^h/(f \cdot p_T^\gamma), Q^2) \langle P(f, p_T^\gamma)\rangle\\
\end{split}
\end{equation}

where we define the fraction of parton energy left after interaction with the medium as $f = \frac{E - \Delta E}{E}$ and transform $\langle P(\Delta E, E)\rangle$ accordingly. For known $D^{vac}(z, Q^2)$, some information about the energy loss distribution can be inferred, dependent on the available momentum range and statistics. We postpone a more detailed analysis of this question to section \ref{S-Gammacorr}.

\section{Sensitivity of $R_{AA}$ to $\langle P(\Delta E, E)\rangle$}

\label{S-P_E}

\begin{figure*}[htb]
\epsfig{file=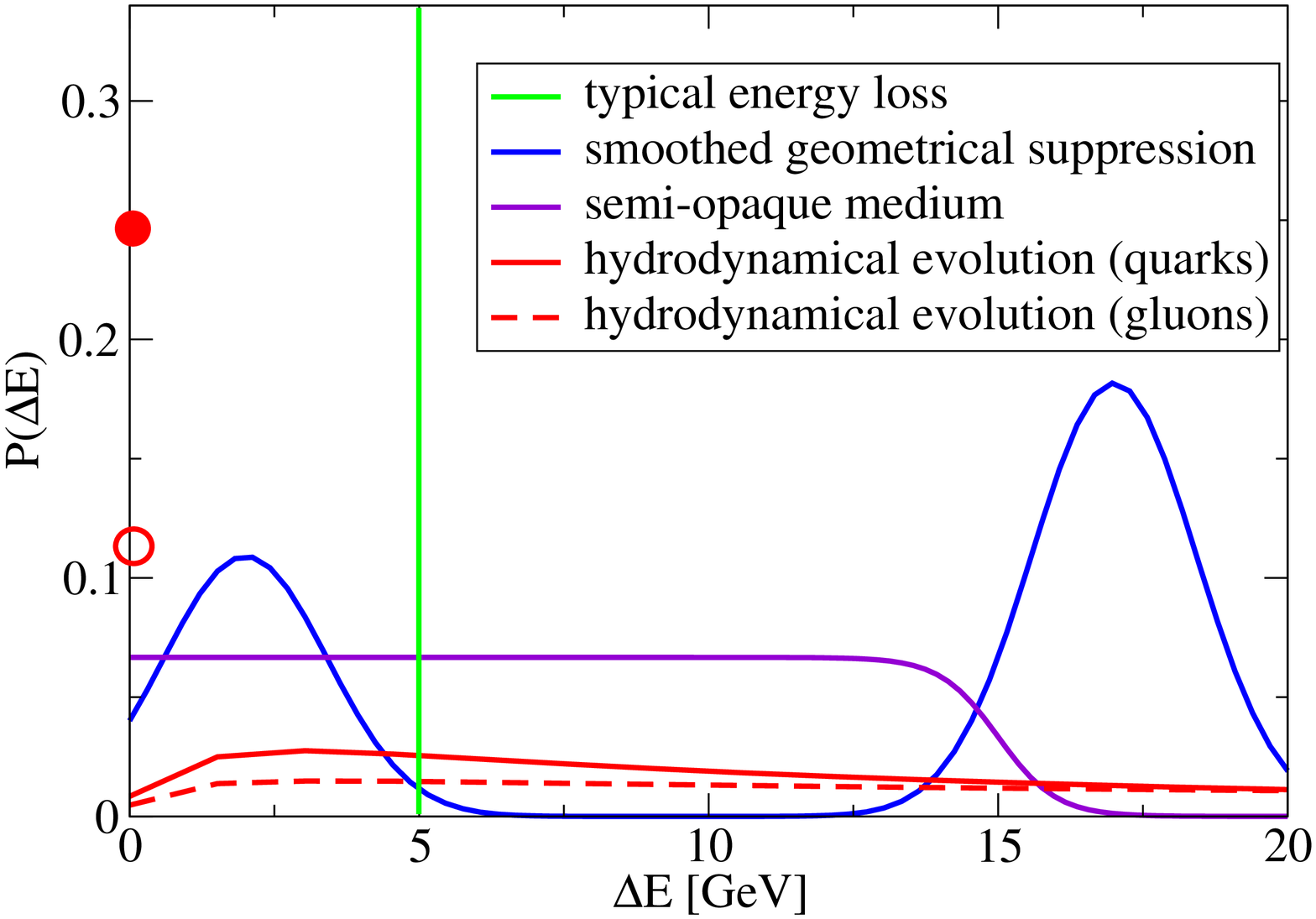, width=8cm} \epsfig{file=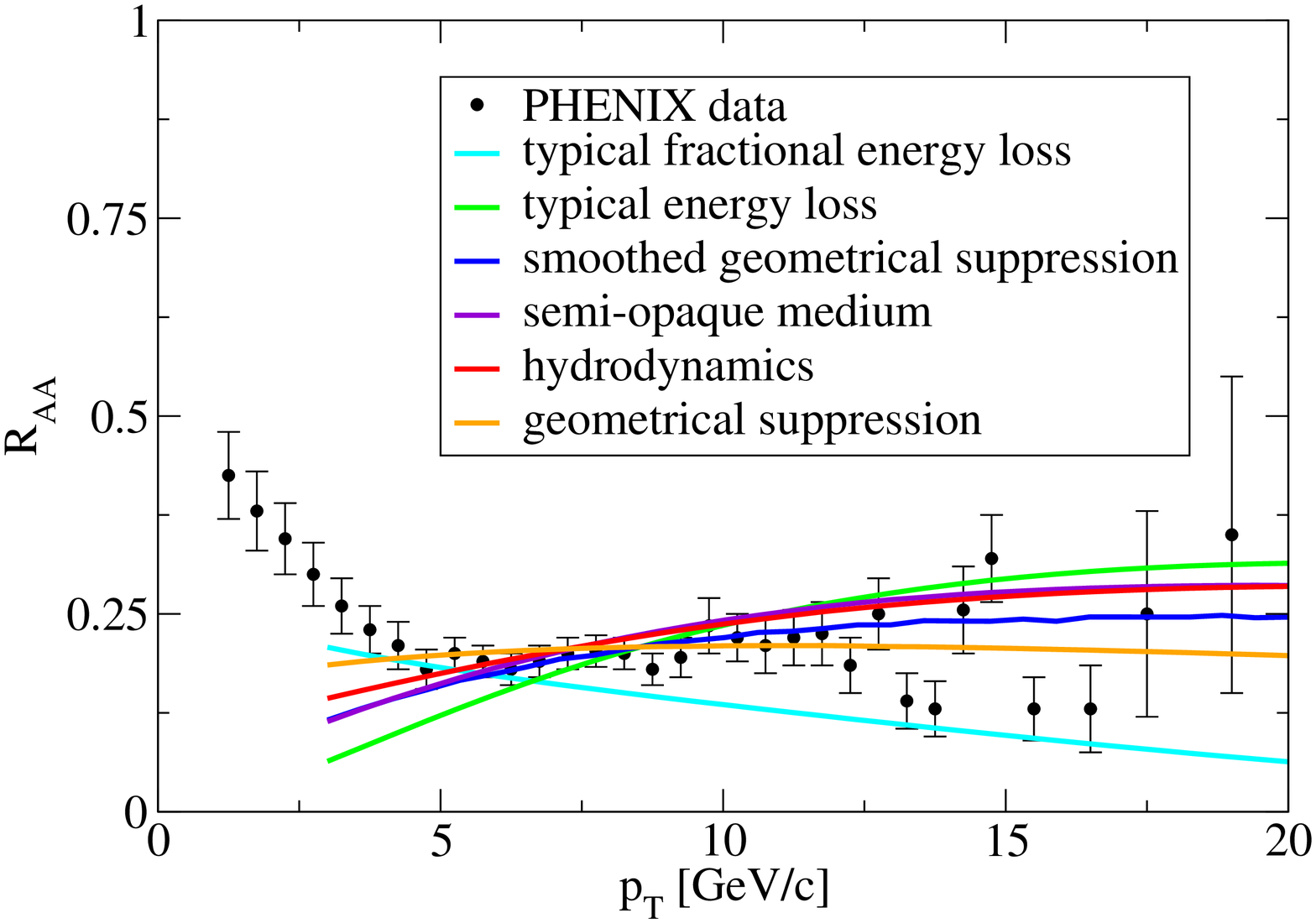, width=8cm}
\caption{\label{F-Comp}Left panel: Averaged energy loss probability distributions for several toy models (see text, energy-dependent distributions are not shown) and a full hydrodynamical simulation. Right: Nuclear suppression factor $R_{AA}$ for different toy models and the hydrodynamical simulation compared with PHENIX data.}
\end{figure*}

Since, after doing the spatial averaging with the medium, $\langle P(\Delta E, E) \rangle$ is the only place in Eq.~(\ref{E-Folding}) where information about the medium enters, from this particular point of view a measurement of $R_{AA}$ only provides constraints for the averaged energy loss probability distribution. The degree of tomographical information in $R_{AA}$ is thus correlated with the sensitivity of $R_{AA}$ to details of $\langle P(\Delta E, E) \rangle$.

There are two main mechanisms which can reduce $R_{AA}$: Absorption of a given fraction of partons ('downward shift' of the spectrum) and energy loss (without absorption) of every parton ('sideward shift' of the spectrum). The true scenario will possibly contain any combination of the two.

To illustrate this point, let us study some simple toy models:

\begin{itemize}
\item Geometrical suppression: We assume that the medium consists of an opaque core and a transparent halo. All partons in the core are absorbed and all partons in the halo escape without energy loss. In this case, $$\langle P(\Delta E, E)\rangle = T \delta(0) + (1-T) \delta(E)$$ (where $T$ is the fraction of transparent region). 

\item Smoothed geometrical suppression: Instead of taking a sharp boundary, we allow for some transition region, i.e. we replace the $\delta$-functions by Gaussians with a width $d$ and replace the infinite opaqueness by some typical energy loss $\Delta E_0$ Then,
  
\begin{displaymath}
\begin{split}
\langle P(\Delta E, E)\rangle &= T \sqrt{\pi} d/2 \exp[-E^2/d^2] \\&+ (1-T) \sqrt{\pi} d \exp[-(E - \Delta E_0)^2/d^2]\\
\end{split}
\end{displaymath}.

\item Typical energy loss: We assume that the whole medium is characterized by a typical energy loss $\Delta E_0$. In this case,  $$\langle P(\Delta E, E)\rangle = \delta(\Delta E_0)$$.

\item Typical fractional energy loss: Here we assume that a hard parton typically loses a constant fraction $(1- f)$ of its energy. Then $$\langle P(\Delta E, E)\rangle = \delta((1-f)E)$$.

\item Extended semi-opaque medium: If the medium is large enough so that all partons are initially inside the medium and transparent enough such that suppression only accumulates for those partons which have a long in-medium pathlength, the suppression probability can be assumed to be flat as a function of $\Delta E$ up to some maximum value $\Delta E_{max}$ (determined by the maximum in-medium pathlength). We model this as $$\langle P(\Delta E, E)\rangle = N/(1 + \exp \left[\frac{\Delta E - \Delta E_{max}}{d}\right])$$ 
with $N$ the normalization factor.

\end{itemize}  

Clearly, the physics (and the medium) underlying these scenarios is very different and any tomographic method should be able to resolve the differences. In Fig.~\ref{F-Comp} we show the different model probability distributions (including one extracted from a hydrodynamic model \cite{Hydro} using the formalism outlined in \cite{JetFlow}). All probability distributions are shown with the convention that a parton is considered to be thermalized and absorbed into the medium as soon as $E - \Delta E = 0.5$ GeV is realized. This determines the fraction of absorbed partons as a function of parton energy from the distributions. We also show the resulting $R_{AA}$ after folding with the parton spectrum and the fragmentation function.

As apparent from Fig.~\ref{F-Comp}, the calculated $R_{AA}$ is for most models well in agreement with the data beyond $p_T = 5$ GeV. In particular, the flatness of $R_{AA}$ is described well in most scenarios. The exception is the constant fractional energy loss where the calculation yield an $R_{AA}$ which continues to drop as a function of $p_T$.  Thus, we may draw as a first conclusion that the flatness of $R_{AA}$ constrains probability distribution to have no strong dependence on the initial parton energy. The actual shape of the distribution is not constrained. Indeed the subset $\langle P(\Delta E)\rangle$ performs rather well.

\section{The magnitude of the quenching}

Naturally, an arbitrary probability distribution does not describe $R_{AA}$ --- in each of the models, one (or at most two) parameters need to be adjusted. For the geometrical suppression, $T=0.18$ is used (note that the resulting curve is {\em not} flat as naively expected as it still reflects the difference between nuclear and proton parton distribution functions), the smoothed geometrical suppression requires $T = 0.3$ and $\Delta E_0 = 17$ GeV (with a weak dependence of the result to $d$), the typical energy loss requires $\Delta E_0$ = 5 GeV, for the typical fractional energy loss we find $f=0.7$ and the extended semi-opaque medium leads to $\Delta E_{max} = 14$ GeV (with a weak dependence on $d$). What is common to all these parameter choices is that they make quenching in some sense substantial, either by setting a large scale for the typical shift of the spectrum or by making the transmission fraction $T$ small. Thus, while the flatness of $R_{AA}$ favours energy independence, its magnitude indicates substantial quenching.

Much effort has been made to condense this qualitative statement into a number, i.e. an averaged transport coefficient $\hat{q}$ which characterizes the typical opaqueness of the early medium. Different groups have reported widely different results, e.g. 5-10 GeV$^2$/fm (Wiedemann/Salgado), 0.35-0.85 GeV$^2$/fm (GLV), $\approx 2$  GeV$^2$/fm (AMY) and 3-4 GeV$^2$/fm (Majumder) \cite{HPTalks}). Based on these results, (dis)-agreement with pQCD expectations has been claimed. 

The main difference in the extraction of $\hat{q}$ in the frameworks cited above seems to be the question how to define the averaging procedure for a medium at which the transport coefficient is in principle different at each point in space and time and only a fraction of the partons may emerge from the medium while others are absorbed (and hence can't carry any information), or even to define a typical average pathlength of a parton in the medium. The same problem is apparent from a different angle in Fig.~\ref{F-Comp} --- why should one expect to be able to extract a typical quenching power of the medium when widely different energy loss probability distributions (corresponding to very different maximal and typical quenching and very different average pathlengths) result in the same $R_{AA}$?

For this reason, we have in \cite{JetFlow} adopted a different procedure: Taking $\hat{q} = 2 \cdot K \cdot \epsilon^{3/4}$ with $K=1$ as the pQCD expectation \cite{Baier}, we can for a given model of the spacetime dynamics fit $K$ to the measured $R_{AA}$ and investigate the disagreement with pQCD by doing the full spatial integration in Eq.~(\ref{E-Folding}) instead of resorting to schematic averaging procedures. We found that, within a class of evolution models which describe bulk hadronic observables such as $m_T$-spectra for $\pi$, K, and p and $dN_{ch}/d\eta$ (which are hence substantially more constrained than typical models used to calculate $R_{AA}$) $K$ can still vary from close to 1 to about 5, dependent on detailed assumptions about the longitudinal and transverse flow profile and the magnitude of $\alpha_s$.

Thus, the conclusion is that even the magnitude of $R_{AA}$ does not allow detailed conclusions about the early evolution dynamics. This, along with the independence of $R_{AA}$ to the detailed shape of the averaged energy loss probability, indicates that there is almost no tomographical information to be gained from $R_{AA}$ alone and that more differential observables need to be studied in order to learn about the early evolution.

\section{Photon-hadron correlations}

\label{S-Gammacorr}

The idea underlying the use of $\gamma$-hadron correlations to distinguish between different scenarios is very intuitive: If we had a monochromatic parton source and could measure the spectrum of outgoing partons, the effects of absorption of a fraction of partons and a sideward shift due to some energy loss would be completely different. In the case of absorption, the intensity of the source would be reduced but it would still be monochromatic, in the case of a shift the intensity would be unchanged but the spectrum would be altered and in general no longer be monochromatic. A photon trigger can provide in principle the necessary monochromatic parton source and a study of the momentum spectrum of hard away side hadrons carries the momentum information. Since the photon escapes from the fireball without further interaction, there is no geometrical trigger bias and the distribution of hard vertices for observed triggers is just given by the nuclear overlap as for $R_{AA}$.

The main practical problem is that the outgoing hadron fragments and we have to solve Eq.~(\ref{E-Analysis}) to gain access to the desired energy loss probability distribution. With the left hand side known at a finite number of points, with finite statistics and in a limited momentum range, this is not an easy task.

We study the problem by embedding $gq \rightarrow \gamma q$ events at a given photon energy of 15 GeV into the Monte-Carlo simulation designed to study dihadron correlations \cite{Dijets}.  By replacing the dynamical energy loss calculation with the toy model probability distributions described in section \ref{S-P_E} (this is possible as the vertex distribution of events is the same as for $R_{AA}$, see above) we can simulate the momentum spectra of hard hadrons back to back with the photon for each of the energy loss models.

In order to extract the energy loss mechanism, we try to solve Eq.~(\ref{E-Analysis}) with a trial ansatz. In the case of geometrical suppression, we can write 
\begin{equation}
\langle P((f, E_\gamma) \rangle = T \cdot \delta(1) + (1-T) \cdot \delta(0) 
\end{equation}

\begin{figure}[htb]
\epsfig{file=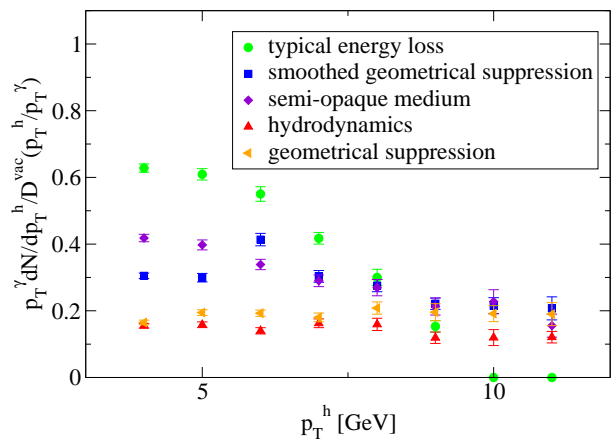, width=8cm}
\caption{\label{F-Photon}Momentum spectrum of hard hadrons correlated back-to-back with a photon trigger, normalized to the expectation of geometrical absorption (see text).}
\end{figure}

Inserting this into Eq.~(\ref{E-Analysis}) yields

\begin{equation}
p_T^\gamma\frac{dN}{dp_T^h} = T \cdot D^{vac}(p_T^h/p_T^\gamma)
\end{equation}

thus, the momentum spectrum divided by the fragmentation function is expected to be flat if geometrical suppression is realized. We show this representation of the simulated data for the different scenarios in Fig.~\ref{F-Photon} (we do not discuss constant fractional energy loss further, as this can be already identified on the level of $R_{AA}$).

The individual models can clearly be told apart. To go beyond that and recover some degree of information about the underlying probability distribution, we make a trial ansatz where we have in addition to the transmittion ('T') term a typical energy loss ('L')

\begin{equation}
\label{E-Trial}
\langle P((f, E_\gamma) \rangle = T \cdot \delta(1) + L \cdot \delta((p_T^\gamma -\Delta E_0)/p_T^\gamma) + (1-T-L) \cdot \delta(0) 
\end{equation}

Using the normalized quantity $p_T^\gamma\frac{dN}{dp_T^h}/ D^{vac}(p_T^h/p_T^\gamma)$, our trial ansatz yields
\begin{equation}
\frac{p_T^\gamma\frac{dN}{dp_T^h}}{D^{vac}\left(\frac{p_T^h}{p_T^\gamma}\right)} = T + L \frac{D^{vac}\left(\frac{p_T^h}{p_T^\gamma - \Delta E_0}\right) }{D^{vac}\left(\frac{p_T^h}{p_T^\gamma}\right)}.
\end{equation}

Clearly, a simultaneous fit of $T,\Delta E_0$ and $L$ is not a promising strategy as the first two terms in Eq.~(\ref{E-Trial}) may become degenerate for $\Delta E_0 \rightarrow 0$. Thus, we chose to adjust $T$ to the lowest data point in each sample and fit $L$ and $\Delta E_0$ accordingly.

The resulting parameters characterize the individual scenarios more or less well. For the geometrical suppression we find $T=0.19$, $L\approx0$ and hence $\Delta E_0$ undefined as expected. For the apparently rather similar hydrodynamics calculation, we find $T=0.12$, $L = 0.07$ and $\Delta E_0 = 4.38$, i.e. in addition to $\sim12$\% partons without energy loss there are some $\sim7$\% partons undergoing on average about 4 GeV energy loss observed, the rest is absorbed.  

Let us next discuss the typical energy loss scenario. Here we find $T=0$ and $L=1.53, \Delta E_0 = 5.24$. While the energy loss value put into the simulation is approximately recovered, the result for $L$ is a surprise as we would expect $T,L <1$ and the sum of transmission, loss and absorption term to be unity. However, due to the finite momentum range considered here, we do not see the full distribution and contributions at low momentum can balance the conservation of probability. 

For the smoothed geometrical suppression the fit yields $T=0.2, L = 0.25$ and $\Delta E_0 = 4.28$. This is rather similar to the results of the semi-opaque medium where we find $T=0.16, L = 0.29$ and $\Delta E_0 = 3.6$. 

Thus, while the ratio of $L/T$ provides some measure to characterize the different scenarios as absorption or loss-dominated, it is instructive that the average value of energy loss 'seen' in this study is always close to 4 GeV. This is presumably rather related to the assumed cuts --- for a 15 GeV parton, significantly more energy loss does not result in much hadronic yield above 4 GeV, significantly less does not result in the correct amount of quenching unless there is an absorption term. Thus, by varying the spread between trigger and associate hadron momentum one opens the possibility to probe more and more of the energy loss probability distribution.

\section{Summary}

We have argued that averaged energy loss probability distributions are a useful tool to compare individual model calculations, as they characterize an energy loss scenario according to the degree of transmission, absorption and energy loss of partons. Making several toy models for different pictures of energy loss, we have demonstrated that $R_{AA}$ is incapable of constraining details of the energy loss probability distribution; a more differential observable is needed instead. Even the average magnitude of the quenching observed is not a useful means to characterize the actual conditions in the medium as it is unclear how the averaging procedure should be defined in a model-independent way. Unfortunately, $R_{AA}$ is, due to its insensitivity to details, unsuitable to provide guidance.

As a possible means to overcome this problem and to access the averaged energy loss probability distributions we have suggested $\gamma$-hadron correlations. We have carried out a proof of concept study and argue that such a measurement is able to make a distinction between the different toy models studied here and also to characterize them to some degree. The same information is in principle accessible via dihadron correlations, cf. \cite{Dijets}. However, in this case the analysis is much complicated by the fact that the initial energy of the partons, given a high $p_T$ trigger hadron, is only known on a probabilistic basis and by the additional problem that energy loss on the trigger leads to a systematic distortion of the vertices leading to a triggered event. While the latter is desireable from the point of view of studying pathlength dependence, it is clearly not advantageous for studying the momentum distribution of the associated hadron in a clean way \cite{Dijets2}.

\begin{acknowledgments}

This work was financially supported by the Academy of Finland, Project 206024.

\end{acknowledgments}

\end{document}